\documentclass[aps,showpacs,nobibnotes, preprint,
nobalancelastpage]{revtex4}
\usepackage{graphicx}
\usepackage{graphics}
\usepackage{amsmath}
\usepackage{amssymb}

\usepackage{latexsym}
\usepackage{dcolumn}
\usepackage{bm}
\newcommand{\simgt}{\lower.5ex\hbox{$\; \buildrel > \over \sim \;$}}
\newcommand{\simlt}{\lower.5ex\hbox{$\; \buildrel < \over \sim \;$}}
\newcommand{\be}{\begin{equation}}
\newcommand{\ee}{\end{equation}}
\newcommand{\ba}{\begin{eqnarray}}

\newcommand{\bi}{\begin{itemize}}
\newcommand{\ei}{\end{itemize}}

\newcommand{\bfi}{\begin{figure}
\epsfxsize=9cm 
\epsffile}
\newcommand{\efi}{\end{figure}}

\newcommand{\mnras}{MNRAS}

\newcommand{\apjs}{ApJS}

\DeclareFontFamily{OT1}{rsfs10}{} 
\DeclareFontShape{OT1}{rsfs10}{m}{n}{ <-> rsfs10 }{} 
\DeclareMathAlphabet{\mathscript}{OT1}{rsfs10}{m}{n} 

\begin{document}

\title{Spherical Collapse in $f(R)$ Gravity}
\author{Alexander Borisov$^1$, Bhuvnesh Jain$^1$, Pengjie Zhang$^2$}
\email{borisov@physics.upenn.edu}
\affiliation{
Department of Physics and Astronomy, University of Pennsylvania,
  Philadelphia, PA 19104\\
}

\begin{abstract}
We use 1-dimensional numerical simulations to study spherical collapse in the $f(R)$ gravity models. We include the nonlinear coupling of the gravitational potential to the scalar field in the theory and use a relaxation scheme to follow the collapse. We find an unusual enhancement in density near the virial radius which may provide observable tests of gravity. We also use the estimated collapse time to calculate the critical overdensity $\delta_c$ used in calculating the mass function and bias of halos. We find that analytical approximations previously used in the literature do not capture the complexity of nonlinear spherical collapse. 
\end{abstract}
\maketitle

\section{Introduction}
The energy contents of the universe pose an interesting puzzle, 
in that general relativity (GR) plus the Standard Model of particle
physics can only account for about $4\%$ of the energy density inferred from
observations.  By introducing dark matter and dark  
energy, which account for the remaining $96\%$ of the total 
energy budget of the universe,  cosmologists have been able to
account for a wide range of  observations, from  
the overall expansion of the universe to 
the large scale structure of the early and late
universe~\cite{Reviews}. The dark matter/dark energy scenario assumes the validity of GR at
galactic and cosmological scales and introduces exotic components of
matter and energy to account for observations. Since 
GR has not been tested independently on these scales, a natural
alternative is that GR itself needs to be modified on large scales.
Two classes of modified gravity (MG) models are higher dimensional scenarios such as the DGP model, and modifications to the Einstein-Hilbert action known as $f(R)$ models~\cite{DGP,fR,Sahni2005}. 
 By design, successful MG models are difficult to distinguishable from viable DE models against observations of
the expansion history  of the universe. However, in general they
predict a different growth of perturbations which can be tested using
observations of large-scale structure (LSS) 
\cite{Yukawa,Stabenau2006,MartSal,Skordis06,Dodelson06,DGPLSS,consistencycheck,Koyama06,fRLSS,Zhang06,Bean06,MMG,Uzan06,Caldwell07,Amendola07, HuSaw3,BoJ}.

Recently the nonlinear regime of structure formation in MG theories has been explored through simulations and analytical studies. Both the DGP and $f(R)$ models have a mechanism that restores the theory to GR on small scales. Recent studies of $f(R)$  theories have focused 
on the effects of the chameleon field which alters the dynamics of mass clustering in high density environments such as galaxy halos. A series of papers ~\cite{HuSaw, Schmidt, OY, HuSaw2, LimaHu} have explored the
consequences of this evolution through simulations and comparison to
analytical predictions. Similar efforts have been made to study large scale structure formation in DGP ~\cite{KhW, Scoc, Sch2,Schmidt2}.

The evolution of isolated spherical overdensities in
an expanding universe provides a useful approximation for structure formation
in the universe.  Although spherical collapse is just
  one  idealized model for the formation of structure, it 
  captures many crucial features of realistic mass distributions. This has been demonstrated by its
successful applications in predicting the halo mass function, halo bias and
merger history in the $\Lambda$CDM cosmology. Spherical collapse is sensitive
to the nature of gravity,  matter and energy.  

Spherical collapse in general
relativity with cold dark matter and smooth dark energy is unique in several
aspects.  For a spherical shell enclosing a fixed mass $M$, its
collapse rate does not rely on the environment nor the internal
density profile. Furthermore, an initial tophat spherical region 
remains a tophat during the collapse. Finally, after virialization, we have the
simple relation $2K+W=0$ between the total kinetic energy $K$ and the
potential energy $W$. All these interesting properties rely on the $r^{-2}$
behavior of the gravitational force. Modifications to general relativity often destroy the 
characteristic properties described above: spherical collapse  can become dependent on 
environment and internal structure, rendering an initial tophat density profile
non-tophat and changing the conversion efficiency from potential energy to
kinetic energy. If gravity indeed deviates from GR, we expect that at least
some of these modifications would survive in realistic galaxies and  galaxy 
clusters and serve as tests of modified gravity. 

Due to the complicated behavior of gravity in $f(R)$ models
 spherical collapse no longer has analytical solutions. The
purpose of this paper is to study spherical collapse
through 1-dimensional numerical simulations.  
Schmidt et al. \cite{Schmidt}  have used large-scale simulations, coupled with analytical 
approximations for spherical 
collapse, to predict the halo mass functions, linear bias and density
profiles for the $f(R)$ model of Hu \& Sawicki \cite{HuSaw} and compared them to the standard
model of cosmology - the  $\Lambda CDM$ model. In addition analytical
calculations have been done in the two limiting cases of the $f(R)$ model: 
\begin{enumerate}
 \item The strong field regime, where $f(R)$ behaves like $\Lambda CDM$, but with a larger Newton's constant (by a factor of $4/3$).
 \item The weak field regime, where there is no observable difference from the Standard Model.
\end{enumerate}
The results from these bounding situations have been compared to simulations by ~\cite{Schmidt} and the observed differences have been discussed. Since the strength of gravity in $f(R)$ gravity lies inside these two limiting cases, a reasonable expectation is that the evolved observable quantities should also lie within the limiting cases. We will explore the validity of such an assumption by performing a direct simulation of a spherical collapse of an isotropic
object. As chameleon $f(R)$ theories exhibit highly non-linear behavior and there exist coupled fields, it is worth checking through explicit calculation the naive expectation based on limiting cases. 

As described in \cite{OY} for simulations of $f(R)$ gravity, the solution for the potential driving the dynamics of the evolution is coupled with the solution for the scalar field $f_R$ \cite{HuSaw}. In our case we deal with isotropic objects and thus have a one-dimensional system. In \S II we present the radial equation for the $f_R$ field. In \S III we describe the simulation scheme for numerically solving the aforementioned equation.  \S IV describes the results, focusing on the distinct features of spherical collapse in $f(R)$ gravity, while \S V connects our results to the mass function. We conclude in \S VI. 

\section{Radial Equations for Spherical Collapse}

In general $f(R)$ models are a modification of the Einstein-Hilbert action of the form:
\begin{equation}
 S = \int d^{4}x \sqrt{-g} \left[ \frac{R + f(R) }{2 \kappa^{2} }  \right], 
\end{equation}
where $R$ is the curvature and $\kappa^{2} = 8 \pi G$. The particular form chosen by ~\cite{HuSaw} is:
\begin{equation}
 f(R) = - m^{2} \frac{c_{1} (R/m^{2})^{n}}{c_{2} (R/m^{2})^{n} + 1} 
\end{equation}
with
\begin{equation}
 m^{2} \equiv \frac{\kappa^{2}\bar{\rho}_{0}}{3} = (8315 Mpc)^{-2} \left( \frac{\Omega_{m} h^{2}}{0.13} \right) 
\end{equation}
 The properties of the model are well described by the auxiliary field $f_{R} \equiv
 \frac{df(R)}{dR}$. 
$\Lambda {\rm CDM}$ expansion history with a cosmological constant $\Omega_{\Lambda}$ is obtained if we set:
\begin{equation}
\frac{c_1}{c_2} \approx 6 \frac{\Omega_{\Lambda}}{\Omega_m} 
\end{equation}
We choose to work with $n=1$, which leaves one free parameter $f_{R0} \approx -n \frac{c_1}{c^2_2} \left(\frac{12}{\Omega_m} -9 \right)^{-n-1}$ to parametrize the model.  Following \cite{OY}, which set up the 3D simulation framework for the $f(R)$ chameleon model, we start with the trace of the Einstein equations in the quasi-static limit, and the Poisson equation:
\begin{equation}
\label{eqn:fr1}
 \nabla^2 f_R = \frac{1}{3 c^2} \left[ \delta R(f_R) - 8 \pi G \delta \rho \right]
\end{equation}
\begin{equation}
\label{eqn:fr2}
 \nabla^2 \phi = \frac{16 \pi G \rho_0}{3} \delta \rho - \frac{1}{6} \delta R(f_R).
\end{equation}
For the purposes of numerical calculations we need to define relevant dimensionless quantities, and we switch to comoving coordinates. Thus we adopt the definition of code units \cite{OY, Shand, Krav}:
\begin{align}
\label{eqn:ccc}
&\tilde r = \frac{x}{r_0 a}, \quad \tilde t = t H_0, \quad \tilde \rho = a^3 \frac{\rho}{\rho_0},
\\ &\tilde R = a^3\frac{R}{R_0}, \quad \tilde c = \frac{c}{r_0 H_0}, \quad \tilde \phi = \frac{\phi}{\phi_0}, \quad \tilde{p}=a \frac{v}{v_0}, \nonumber
\end{align}
 where
\begin{equation}
\label{eqn:ccce}
 \rho_0 = \rho_{c,0}\Omega_{M,0}, \quad R_0 = \frac{8 \pi G \rho_0}{3}, \quad \phi_0 = (r_0 H_0)^2, \quad v_0 = r_0 H_0
\end{equation}
and $r_0$ is an appropriate length scale (used for example to define the size of the overdensity). Bare symbols $X$ are physical coordinates/quantities while symbols with tilde $\tilde X$ are code quantities, symbols with bars $\bar X$ are average physical and symbols with both bar and tilde $\bar{\tilde X}$ are average code quantities.

Equations (\ref{eqn:fr1} and \ref{eqn:fr2}) then become the following in code units (Eqns. 25,27 in \cite{OY}):
\begin{equation}
 \tilde \nabla^2 \delta f_R = \frac{\Omega_{M,0}}{a \tilde c^2} \left[ \frac{\delta \tilde R}{3} - \delta \right]
\end{equation}
\begin{equation}
 \tilde \nabla^2 \tilde \phi = \frac{\Omega_{M,0}}{a} \left[ - \frac{\delta \tilde R}{6} + 2 \delta \right]
\end{equation}
where 
\begin{equation}
 \delta = \frac{\rho - \bar \rho}{\bar \rho} = \tilde \delta.
\end{equation}
The next step is to express $\delta \tilde R$ in terms of $f_R$ in code
coordinates. 
We start with (Eqn 9 in \cite{OY}):
\begin{equation}
\label{eqn:strf}
 \bar{R} =  8 \pi G \bar{\rho}_M \left( \frac{1}{a^3} +4 \frac{\Omega_{\Lambda,0}}{\Omega_{M,0}} \right)
\end{equation}

This is the average curvature in the $f(R)$ model.
Thus:
\begin{equation}
 \frac{\bar{R}}{R_0} = 3\frac{\bar{\rho}_M}{\rho_0}\left( \frac{1}{a^3} +4 \frac{\Omega_{\Lambda,0}}{\Omega_{M,0}} \right)
\end{equation}
We arrive at:
\begin{equation}
 \frac{\bar{R}}{R_0} = 3\left( \frac{1}{a^3} +4 \frac{\Omega_{\Lambda,0}}{\Omega_{M,0}} \right)
\end{equation}
From that we also have:
\begin{equation}
 \frac{\bar{R}(a=1)}{R_0} = 3\left(1 +4 \frac{\Omega_{\Lambda,0}}{\Omega_{M,0}} \right)
\end{equation}
We also need the relation between $f_R$ and $R$. Using (Eqn. 12 in ~\cite{OY}), defining $\bar{f}_R(a=1) = f_{R0}$, and working in the case $n=1$ we see that:
\begin{equation}
 \frac{f_R}{f_{R0}} = \left(\frac{\bar{R}(a=1)}{R} \right)^2
\end{equation}
This leads to:
\begin{equation}
 \frac{R}{R_0}=\frac{R}{\bar{R}(a=1)} \frac{\bar{R}(a=1)}{R_0}= 3\left(1 +4 \frac{\Omega_{\Lambda,0}}{\Omega_{M,0}} \right) \sqrt{\frac{f_{R0}}{f_R}}
\end{equation}
For further use we will also need:
\begin{equation}
\sqrt{ \frac{f_{R0}}{\bar{f}_R}} = \frac{\bar R}{\bar{R}(a=1)} = \frac{\left( \frac{1}{a^3} +4 \frac{\Omega_{\Lambda,0}}{\Omega_{M,0}} \right)}{\left(1 +4 \frac{\Omega_{\Lambda,0}}{\Omega_{M,0}} \right)}
\end{equation}
We can now obtain for the perturbation in the Ricci scalar:
\begin{align}
 &\delta \tilde R = \tilde R - \bar{\tilde R}= \frac{a^3}{R_0} (R - \bar R ) = \frac{a^3}{R_0} \delta R = \\ &= 3 a^3 \left(1 +4 \frac{\Omega_{\Lambda,0}}{\Omega_{M,0}} \right) \left[\sqrt{\frac{f_{R0}}{f_R}} - \frac{\left( \frac{1}{a^3} +4 \frac{\Omega_{\Lambda,0}}{\Omega_{M,0}} \right)}{\left(1 +4 \frac{\Omega_{\Lambda,0}}{\Omega_{M,0}} \right)} \right] \nonumber
\end{align}
From now on we will use only code quantities and drop the tilde notation.
Noting that 
\begin{equation}
 \frac{1}{r^2}\frac{\partial}{\partial r}\left( r^2 \frac{\partial X}{\partial r} \right) =  \frac{1}{r} \frac{\partial^2 (r X)}{\partial r^2}
\end{equation}
 let's consider the following substitution:
\begin{equation}
 f_R = \frac{\bar{f}_R e^{u}}{r}
\end{equation}
Expanding the Laplacian (in code units) we arrive at the following equation for $f_R$:
\begin{equation}
 \bar{f}_R   \frac{1}{r} \frac{\partial^2 }{\partial r^2} e^u =  \frac{\Omega_{M,0}}{a \tilde c^2} \left[ a^3 \left( \frac{1}{a^3} +4 \frac{\Omega_{\Lambda,0}}{\Omega_{M,0}} \right) \left(\sqrt{r}e^{-u/2} -1 \right) - \delta \right]
\end{equation}
Additionally we will convert it to a system of 2 first order ODEs using an auxiliary function:
\begin{equation}
 y=\frac{\partial}{\partial r} e^u = e^u \frac{\partial}{\partial r} u = e^u u'
\end{equation}
So the system with explicit $r$ dependence looks like:
\begin{equation}
 u'(r) = e^{-u(r)}y(r)
\label{eqn:SODE1}
\end{equation}
\begin{equation}
 y'(r) =\frac{r}{\bar{f}_R} \frac{\Omega_{M,0}}{a \tilde c^2} \left[ a^3 \left( \frac{1}{a^3} +4 \frac{\Omega_{\Lambda,0}}{\Omega_{M,0}} \right) \left(\sqrt{r}e^{-u(r)/2} -1 \right) - \delta (r) \right]
\label{eqn:SODE2}
\end{equation}
A detailed description of the relaxation method used to solve this system is presented in Appendix A.

Considering that we can approximate $r_{fi} \sim r_{\infty}$, where $r_{fi}$ is the upper boundary of integration in code coordinates, we can impose the boundary condition that $f_R(r_{fi}) = \bar{f}_R$ which translates to $u(r_{fi})={\rm ln} (r_{fi})$. As the relaxation scheme requires 2 boundary points we will impose a condition on the inner boundary. We do not know the solution in the center of a collapsing isotropic mass distribution. What we know is that it has to be symmetrical. We also expect the solution in the center to be screened
from the solution outside of the sphere by the thin screen (that is we expect to
have chameleon effect). This means that very close to the center we expect to
have behavior very similar to that of a homogeneous universe with that average
density - but that would be a constant solution and thus zero derivative $f_R'(0)=0$. 



\section{Simulation Scheme}

To obtain the time evolution in the simulation, at each time step we proceed as follows:
\begin{enumerate}
 \item Given an initial density profile (from the previous time step) we compute the corresponding solution for the $f_R$ field. Under the quasi-static approximation that we adopt, the gravity field is completely determined by the density distribution at the same epoch. 
  \item This allows us to compute the solution for the Newtonian potential that drives the dynamics.
 \item The mass shells are then moved according to the dynamics equations ~\cite{OY}:

\begin{equation}
\frac{d\tilde{r}}{da} = \frac{\tilde{p}}{\dot{a}a^2} 
\end{equation}
and 
\begin{equation}
\frac{d\tilde{p}}{da} = -\frac{\nabla \tilde{\phi}}{\dot{a}},
\end{equation}
where $\dot{a}=a^{-1/2} \sqrt{\Omega_{M,0} + \Omega_{\Lambda,0} a^3 }$, as we tune the expansion of the universe to be the same as in $\Lambda CDM$.
 \item After the particles (shells) have been moved we can compute the new density distribution and proceed to a new time step thus closing the cycle.
\end{enumerate}

\subsection{Code Tests}

\begin{figure}[hbtp]
\renewcommand{\baselinestretch}{1}
\begin{center}
\leavevmode

\begin{minipage}[b]{1\textwidth}
\includegraphics[width=0.9\textwidth]{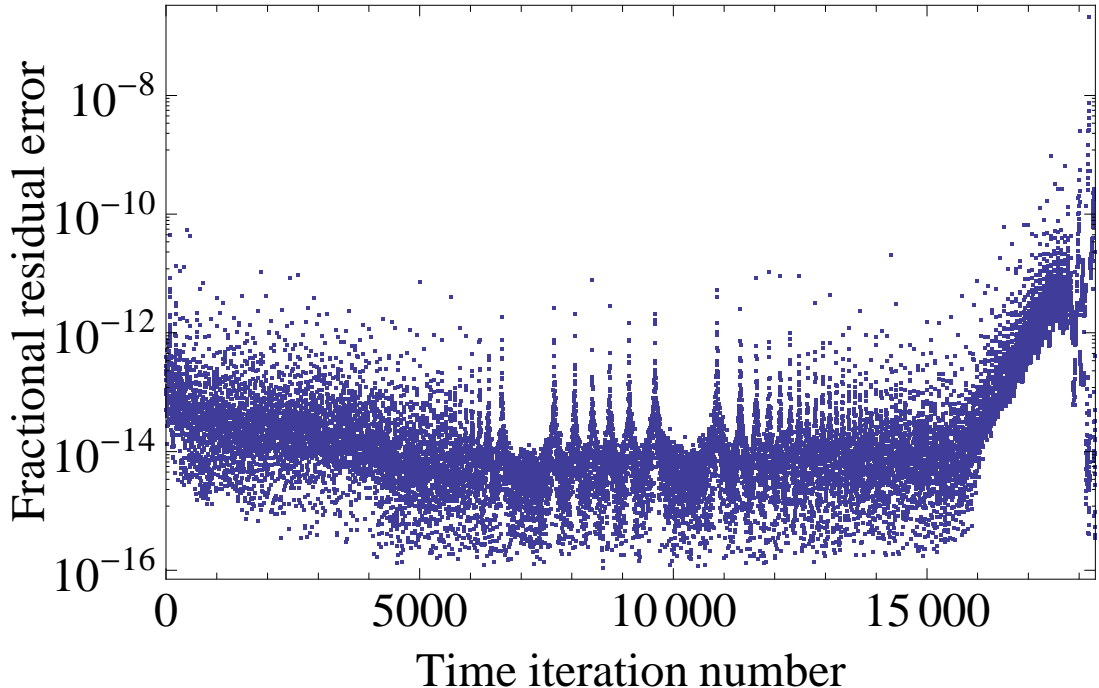}
\end{minipage}%

\end{center}

\caption{\label{III:fig:fig1}{\footnotesize (Color online) Fractional residuals in the solution for $f_R$ at the final step of the relaxation process as a function of the time step}. }
\end{figure}

An important issue in the case of numerical simulations is testing the code for stability and accuracy. The following have been checked:
\begin{enumerate}
 \item Self consistency: as per \cite{OY} we can start with an analytical function for $f_R(r)$. This can be analytically solved to obtain a corresponding density distribution. Now we can plug that density distribution in the numerical code and check how well the obtained solution reproduces the original analytical function. We observe deviations of the order less than $10^{-7}$. 
 \item During the relaxation scheme a measure of our accuracy is the residual relative size of the elements in the vector $\mathbf b$ (Eq.\ref{eqn:matrix}) as compared to the size of the corresponding elements of the solution $\frac{\Delta {\mathbf y}_k}{{\mathbf y}_k}$.  This tells us how much we need to correct the solution obtained by the previous step in the relaxation. A sample plot of these as a function of time iteration step is provided at Fig.\ref{III:fig:fig1}. 
 \item We have also checked the simulation for analytically solvable dynamics. In particular we observe that for simulations of 20000 time iterations we recover the analytical results (turn around radius, virial radius, collapse time) for the Einstein-De Sitter (matter dominated $\Omega_m = 1$) model within 0.3\%. This is the dominant numerical error. 
\end{enumerate}

\section{Virialization in $f(R)$ gravity}

An important question we need to address  is how to identify the epoch at which a collapsing object reaches its virial radius. In the cases of Einstein-De Sitter ($\Omega_{M,0} = 1$) and $\Lambda CDM$ universes we have analytic solutions ~\cite{Schmidt} (Appendix A), but that is not so in the case of $f(R)$ gravity. What we do employ is a step by step calculation of the Virial condition. Let's look at energy conservation. The total velocity $v^t=dx/dt=d(ar)/dt=\dot{a}r+v$,
where $x=ar$ is the physical distance, $r$ is the comoving distance and 
$v=a\dot{r}$ is the proper peculiar velocity. The acceleration equation is
\be
\frac{d(av_p)}{dt}=-\frac{d\phi}{dr}
\ee
On the other hand, $v^t$ satisfies another equation
\be
\label{eqn:vt}
\dot{v^t}=-\frac{d\phi^t}{dx}\ \ ;\ \ \phi^t=\phi-\frac{1}{2}a\ddot{a}r^2
\ee
Multiplying $v^t$ to both sides and integrating over $t$, we obtain the familiar energy conservation
\be
\frac{1}{2}(v^t)^2+\phi^t=constant
\ee
For this reason, $v^t$ and $\phi^t$ are the relevant quantities for the virial
theorem.  Multiply Eq. \ref{eqn:vt} by $x$, we obtain
\be
\frac{d}{dt}(xv^t)-(v^t)^2=-x\frac{d\phi^t}{dx}
\ee
This equation is satisfied at all time. After virialization, we then take the
average of the above equation for all particles. Now the velocity of
particles is random (no correlation with $x$), so we have $\langle
xv^t\rangle=0$. Then 
\be
\langle (v^t)^2\rangle=\langle x\frac{d\phi^t}{dx}\rangle
\ee
An equivalent expression, which can be applied straightforwardly, is 
\be
2K\equiv \int  (v^t)^2 dM=\int x\frac{d\phi^t}{dx} dM \ .
\ee
Here, $K$ is the total kinetic energy. The integral is over the region of mass
$M$. This is the general expression of the virial theorem. One can check in the case of
Newtonian gravity, $\phi^t\propto M/x$ and $\int xd\phi^t/dx dM=-\int \phi^t
dM\equiv -W$, this reduces to familiar form of the
virial theorem, $2K+W=0$. See discussion in \cite{Schmidt2} for the pitfalls of using the Newtonian  potential energy rather than the RHS of Eqn. 29 for modified gravity or general quintessence models. For the purposes of the simulation we need to express the above formulae in terms of comoving code coordinates given by Eqs. \ref{eqn:ccc} and \ref{eqn:ccce}. It is straightforwards to obtain:

\begin{equation}
 K = r_0^2 \int dM \left( \tilde{r} \dot{a} + \frac{H_0 \tilde{p}}{a} \right)^2
\end{equation}
for the kinetic term, and:
\begin{equation}
 W = r_0^2 \int dM \tilde{r} \left( H_0^2 \frac{d \tilde{\phi}}{d \tilde{r}} -\tilde{r} a \ddot{a} \right)
\end{equation}
for the potential term. These are subsequently discretized and the sum is over the region with relevant mass.

As expected in the case of GR (EDS and $\Lambda CDM$) the epoch at which the sum of these two terms is zero coincides with the analytically predicted epoch of reaching the virial radius \cite{Schmidt, Lahav}:
\begin{align}
 \eta = \frac{\rho_{{\rm eff}}}{(1+F)\rho_m} &= \frac{2 \Omega_{\Lambda}}{(1+F)\Omega_m a^{-3} (1+\delta)} \\ \eta &= \frac{2s-1}{2s^3 -1}, \nonumber
\end{align}
where $s=r_v/r_{TA}$ is the ratio of the virial radius and the maximal radius at turn-around. All relevant quantities are defined at turn-around.

\begin{figure}[hbtp]
\renewcommand{\baselinestretch}{1}
\begin{center}
\leavevmode

\begin{minipage}[b]{1\textwidth}
\includegraphics[width=0.9\textwidth]{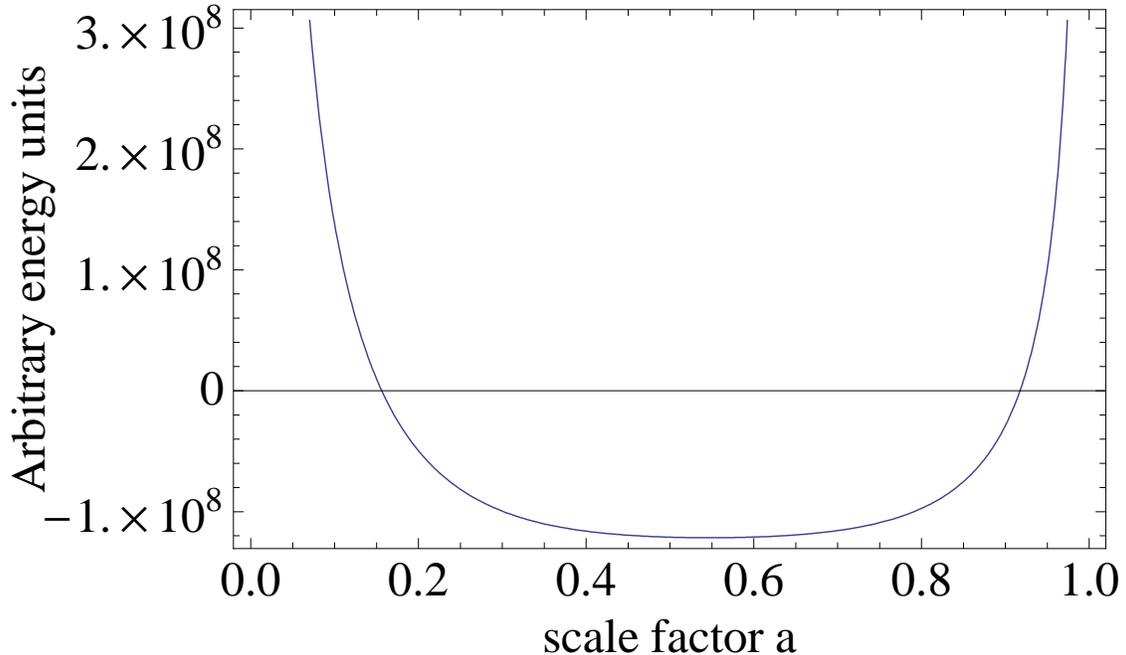}
\end{minipage}%

\end{center}

\caption{\label{III:fig:fig2}{\footnotesize (Color online) Evolution of the virial term. Observe that there are two points where it crosses zero. We are interested in the one that happens after turnaround.} }
\end{figure}

In our simulations we observe that the difference between the analytical result and our evaluation is of order $10^{-5}$ for 20,000 time steps. We expect a similar level of accuracy to hold in the case of $f(R)$ modifications. Thus we define the epoch of achieving virial radius in $f(R)$ by the moment when this sum becomes zero during simulations. As per Fig. \ref{III:fig:fig2} observe that there are two moments when this condition is satisfied. We are obviously interested in the second one, which occurs after passing the turnaround point.

\subsection{Density Enhancement at the Virial Radius}

For the study of spherical collapse in ordinary GR an 
initial top-hat density distributions is very convenient as it 
remains a top-hat during collapse. (In other words a top-hat is a Green's
function for the spherical collapse evolution operator in GR). This allows for
a straightforward definition of the key variables for spherical collapse
- in particular $\delta_c$ and $\Delta_{vir}$. This is not the case for modified
gravity theories. Unfortunately we do not know what profile would be the
analogous Green's function. We can still  compute  the evolution of an initial 
top-hat distribution and try to compute $\delta_c$ and $\Delta_{vir}$ in a similar
way. 

We find that at the outer edge of
the initial distribution the density becomes very large. This can lead to observable signatures
for cluster halos, e.g. in weak lensing mass profiles. 
This effect appears qualitatively consistent 
with arising from chameleon screening. 
As the universe expands the size of the background $f_R$ field
increases and we approach the high field limit of the $f(R)$ theory - where it
behaves as GR with enhanced Newton's constant. This means that the outer edge
 collapses faster than it would do in regular GR. The inside of the
collapsing object, though, is under the effect of the chameleon and the
solution for the $f_R$ field becomes much smaller and thus the collapse slows
down to approach the one in GR with  Newton's
constant. This makes the edge more and more dense as compared to the
inside of the object, and this creates a positive feedback. The higher the
edge density the stronger its screening effect and the inside slows down even
further thus enhancing the accumulation of matter at the edge. However it has  
been suggested (Fabian Schmidt, private discussion) that the density enhancement 
we observe occurs for $f(R)$ models with parameters that do not involve 
chameleon screening. A detailed study of environment and the nonlinearity of 
the theory is required to address the origin of the density enhancement. 

There is an interesting possibility (Justin Khoury, private communication): 
this density enhancement at the edge can
separate the inside and the outside with an underdensity. We actually observe that the
solution for $f_R$ starts exhibiting that kind of behavior in the very late
stages of the collapse, but it is very close to the epoch of reaching virial
radius so the effect is not observable in the density profiles. 
Clearly there are several issues in the late stages
of spherical collapse that merit further study.  
\begin{figure}[thbp]
\renewcommand{\baselinestretch}{1}
\begin{center}
\leavevmode
\begin{minipage}[b]{1\textwidth}
\includegraphics[width=0.9\textwidth]{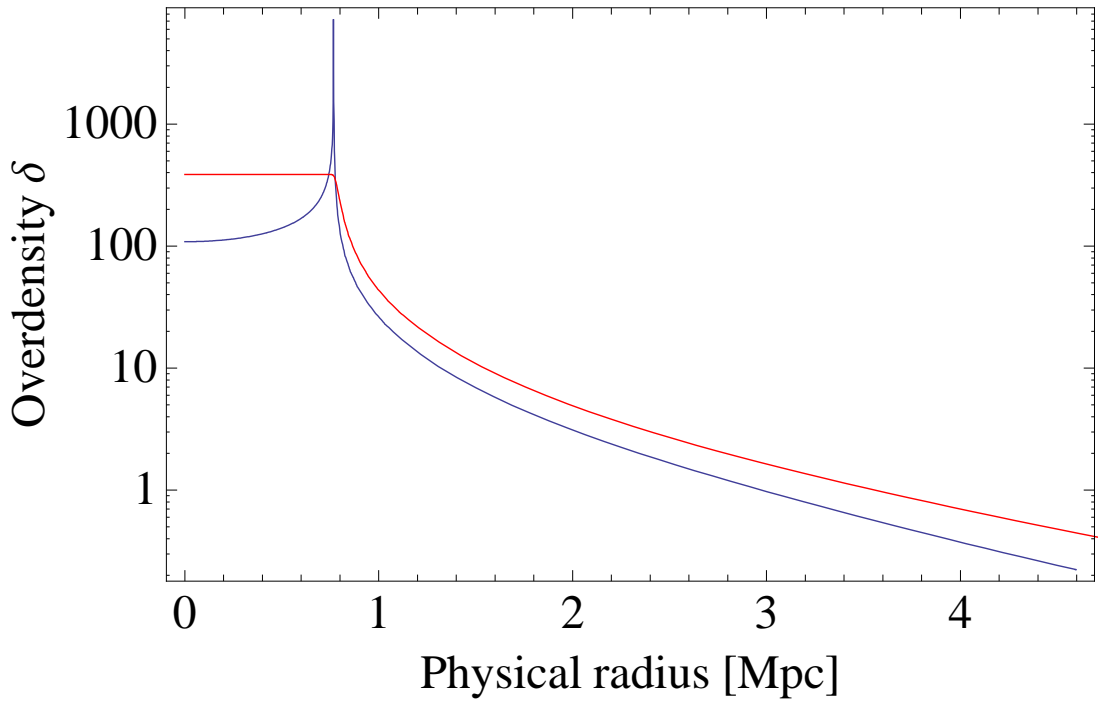}
\end{minipage}%
\end{center}
\caption{\label{III:fig:fig2A}{\footnotesize (Color online) Comparison of the density profiles at the epoch of virialization for  $f(R)$ gravity (blue) and GR (red). In each case the starting profile (Gaussian smoothed tophat) and mass ($1.5 \times 10^{14} \, {\rm M}_{\odot}$) are the same. Virialization is reached at different epochs.} }
\end{figure}

\begin{figure}[htbp]
\renewcommand{\baselinestretch}{1}
\begin{center}
\leavevmode
\begin{minipage}[b]{1\textwidth}
\includegraphics[width=0.9\textwidth]{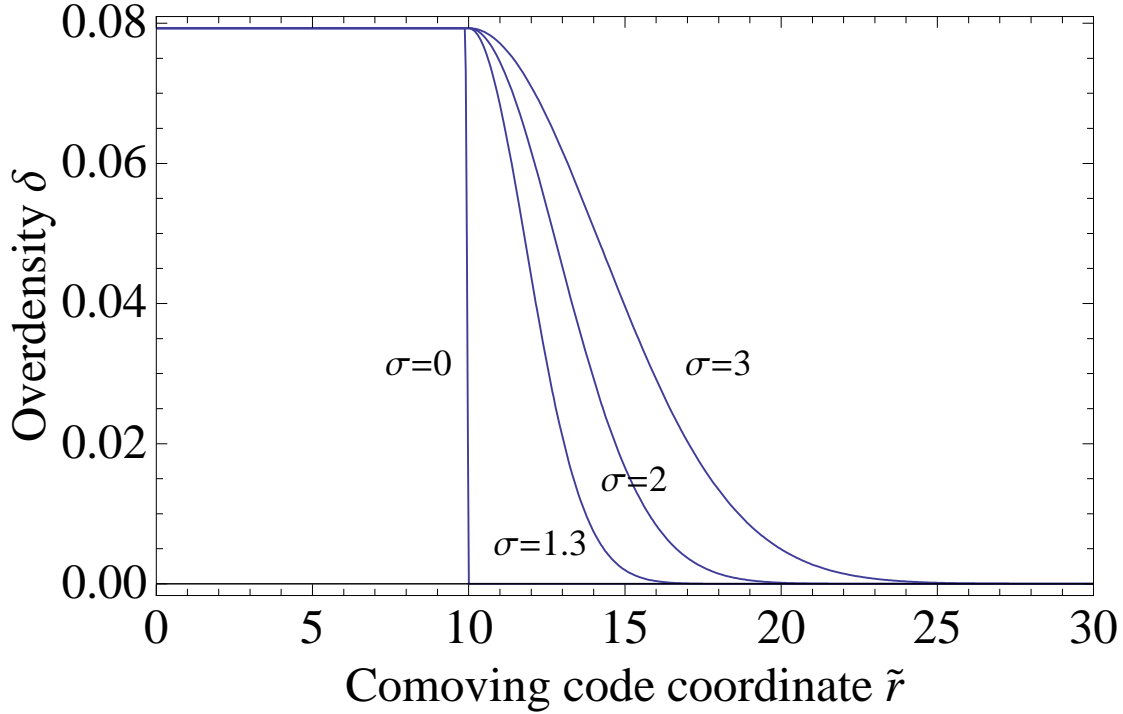}
\end{minipage}%
\end{center}
\caption{\label{III:fig:fig4}{\footnotesize (Color online)  Smoothed density profiles with Gaussians with different dispersion}. }
\end{figure}

The edge effect leads to difficulties in the numerical integration for the initial top-hat. 
We therefore smooth out the edge with a Gaussian, 
which reduces the strength of the positive feedback and
allows for stable evolution of the collapse. This smoothing is applied only 
to the original profile (step 1 of the simulation). The complicated part of
that approach is that, as Birkhoff's theorem is not satisfied in $f(R)$ models
of gravity, the end result significantly depends on the environment, and in
particular, what smoothing is used. 

The smoothed profile has one parameter, 
the dispersion of the Gaussian, which allows for controlling how close
we are to a pure top-hat density distribution. The profile is:
\begin{equation}
 \delta (r) = \delta_{in} ({\rm He}(r) - {\rm He}(r - r_{TH})) + \delta_{in} {\rm He}(r - r_{TH}) e^{-(r - r_{TH})^2/(2\sigma)^2},
\end{equation}
where ${\rm He}$ is the Heavyside step function.

 Even with smoothing, though,
the code becomes unstable beyond reaching the epoch of the virial radius,
preventing us from achieving collapse to singularity - the epoch of
collapse. In Fig. \ref{III:fig:fig2A} we show a comparison of the density
profiles at virialization between $f(R)$ and $\Lambda CDM$. In each case the
starting profile and mass ($1.5 \times 10^{14} \, {\rm M}_{\odot}$) are the same. They achieve virialization at
different epochs. 

\begin{figure}[thbp]
\renewcommand{\baselinestretch}{1}
\begin{center}
\leavevmode

\begin{minipage}[b]{1\textwidth}
\includegraphics[width=0.9\textwidth]{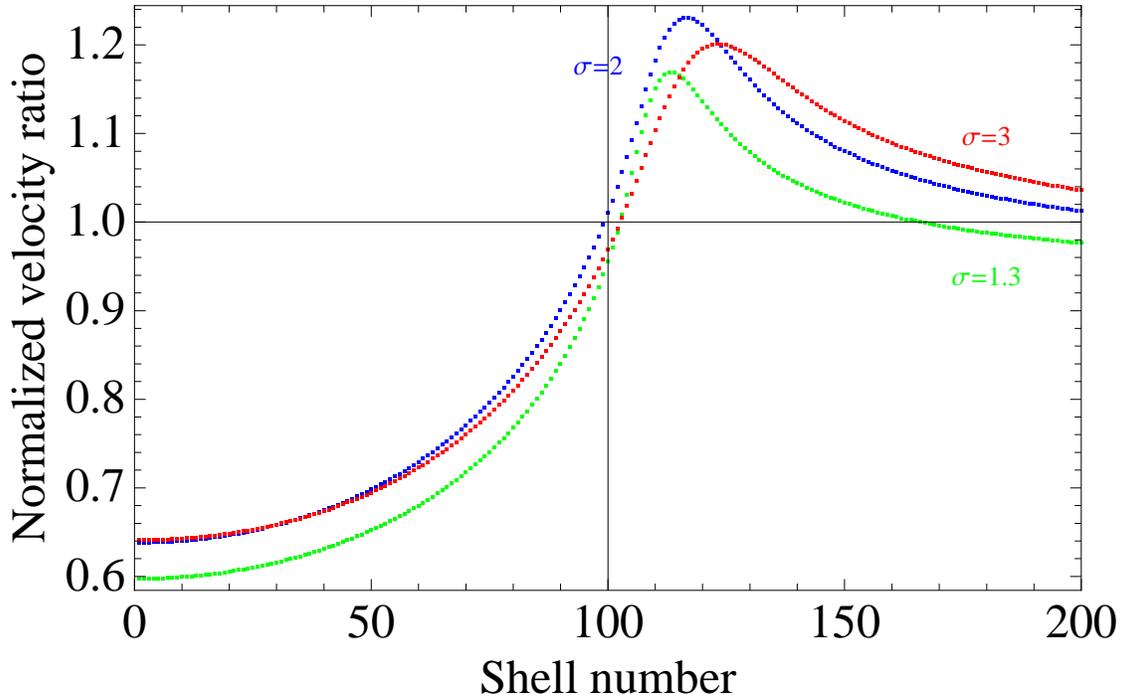}
\end{minipage}%

\end{center}

\caption{\label{III:fig:fig2B}{\footnotesize (Color online) Velocity
ratio computed shell by shell and normalized by the physical position of the
shells. Shell number 100 represents the edge of the top-hat part of the initial
overdensity}. }
\end{figure}

\section{Implications for the mass function: the collapse threshold $\delta_c$}


 \cite{Schmidt} dealt with spherical collapse
 in an analytical way (Appendix A) by solving the 2 limiting cases for the strength
 of the effective Newton's constant in the $f(R)$ model of gravity. The
 prediction in the end is that the fundamental quantities should lie inside of
 the region bound by the values of the 2 limiting cases. In particular they
 are identified by the value of the parameter $F$ which governs the strength
 of the effective Newton's constant. Regular GR corresponds to $F=0$ and the
 strong field limit to $F=1/3$. For $\Omega_{M,0}=0.24$ these imply
 $\delta_c=1.673$ for $F=0$, and $\delta_c=1.692$ for $F=1/3$.

Computing $\delta_c$ is straightforward in $\Lambda CDM$ with GR. For a given starting epoch
$a_{in}$ we need to find an initial overdensity $\delta_{in,GR}$, which would
collapse to a singularity at the present time. Then we just need to evolve that
initial overdensity to present time via the linear growth factor. In the case
of $\Lambda CDM$ this can be performed analytically (\cite{GRcollapse}
Appendix A). In the case of $f(R)$ modified gravity there are 2 complications. 
\begin{enumerate}
 \item The linear growth factor is scale dependent.
 \item Our simulation allows us to only reach the epoch of reaching the virial radius and not the epoch of collapse.
\end{enumerate}

Resolving the first issue is not a complicated task. The solution is to go to
Fourier space and convolve the linear growth factor at the epoch of collapse
(normalized with the growth factor at the initial epoch) with the Fourier
image of a top-hat function. After that, we need to Fourier transform back to
physical space, which is greatly simplified, as we are interested only at the
value at $r=0$, and sums up to the evaluation of an integral.

\begin{figure}[thbp]
\renewcommand{\baselinestretch}{1}
\begin{center}
\leavevmode

\begin{minipage}[b]{1\textwidth}
\includegraphics[width=0.9\textwidth]{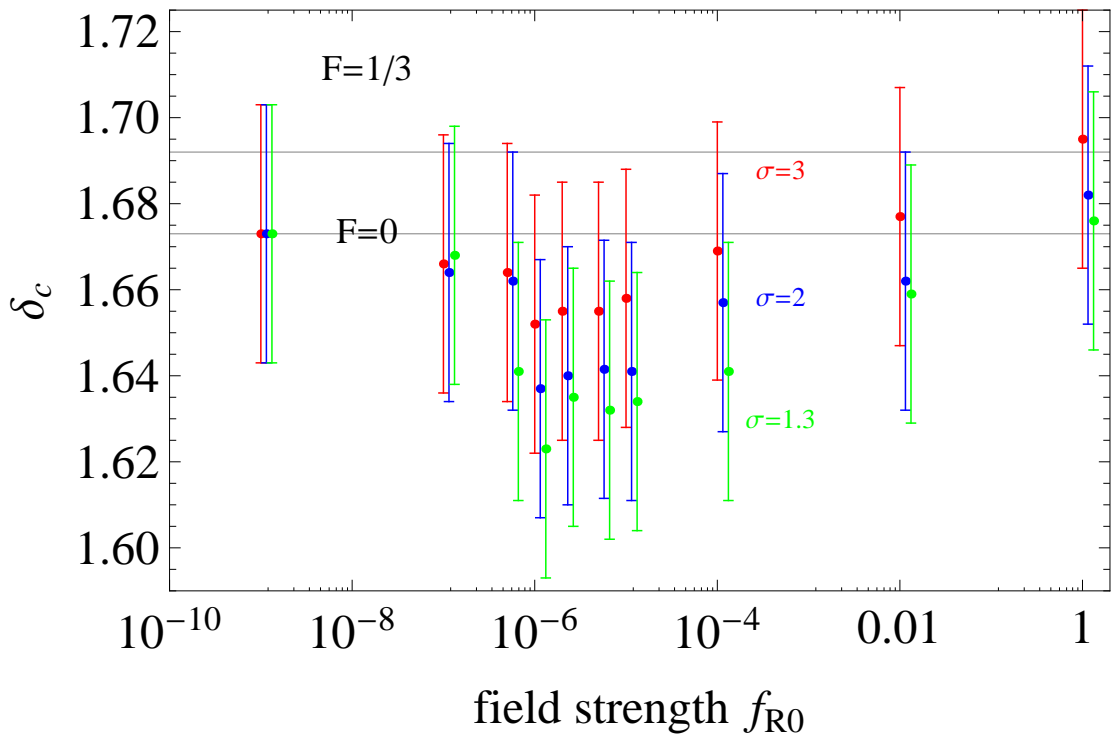}
\end{minipage}

\end{center}

\caption{\label{III:fig:figdc}{\footnotesize  (Color online) $\delta_c$ as a function of field strength $f_{R0}$ for mass ($1.5 \times 10^{14} \, {\rm M}_{\odot}$). The results are color coded to represent the dispersion of the smoothing Gaussian.} }
\end{figure}

Next we deal with the problem of estimating the collapse epoch. 
The numerical issues we have with the development of a density spike at the edge
require the use of approximations. First 
we  study the effect of the environment due to the invalidity of Birkhoff's
theorem. Recall that if Birkhoff's theorem is valid in a gravitational theory
then the behavior of a shell depends only on the mass inside the ball enclosed
by that shell. In our case that is not correct and shells are influenced by
what is outside -- the environment.    
We utilize a sequence of initial profiles, each of which has a pure top-hat
part and then is smoothed with a Gaussian with varying dispersions
Fig.\ref{III:fig:fig4} that bring us closer and closer to a pure top-hat
distribution. This way we can study the trend of changes in the top-hat part
of the initial profile when approaching a pure top-hat overdensity.  


In order to estimate the collapse epoch, we first note 
that the time/scale factor between the epoch of reaching the virial radius and
the epoch of collapse to singularity is a small fraction of the total
time/scale factor in the evolution of the spherical object. Thus we will
assume that if an object in $f(R)$ gravity achieves its virial radius at the
same epoch as a corresponding object in $\Lambda CDM$ does, then these two
objects should reach collapse to a singularity at approximately the same epoch
as well. Then we can make an estimate of how wrong we are in this
prediction. The task of finding $\delta_c$ then is moved to finding the initial
overdensity in $f(R)$, which would reach its virial radius at the same epoch
at which a corresponding object in GR does. In addition we require that the GR
object collapses to a singularity at the present epoch. 
 
What is left is estimating the error of this calculation. One way to approach
the problem is to look at the radial velocity field of the evolving shells
in our simulation and compare them between the corresponding objects in
$f(R)$ and $\Lambda CDM$. In Fig. \ref{III:fig:fig2B} we show the velocity
ratio computed shell by shell and normalized by the physical position of the
shells (the corresponding $f(R)$ and $\Lambda CDM$ have different mass and
different size at the epoch of achieving the virial radius). The different
colors correspond to the different smoothing factors we have introduced as way
to approach a pure top-hat distribution. In our simulation we have chosen
shell number 100 to represent the edge of the top-hat part of the initial
overdensity. As we can see the normalized velocity ratio remains within 5\% of
unity at the edge of the top-hat, 
which suggests that a good estimate of our error would be of the same order. 

Another way to approach the issue is to vary the initial overdensity and look at how much it changes the epoch of achieving the virial radius and compare with the expected epoch of collapse. In particular, values for the initial overdensity, that have epoch of virial radius close or beyond the expected epoch of collapse, set a bound on our error. We found that this also puts a hard error bar of 5\%, which is what we finally used in our calculation.


\section{Results and discussion}

The main results of this work is presented in Figs. \ref{III:fig:fig2A} and \ref{III:fig:figdc}. The development of an excess overdensity at the edge of spherical halos in $f(R)$ gravity is shown in 
Figs. \ref{III:fig:fig2A}. While we have not carried out detailed studies of realistic mass profiles, the results suggest that the region around the virial radii of cluster halos may contain signatures 
of $f(R)-$type theories of gravity. 

For the collapse threshold $\delta_c$, our results are shown in Fig. \ref{III:fig:figdc}. We tested the following conjectures:
\begin{enumerate}
 \item In the weak field regime our calculations should approach the result for regular strength $\Lambda CDM$ ($F=0$).
 \item In the strong field regime the result must approach the values predicted in \cite{Schmidt} for $F=1/3$. This behavior is not guaranteed. We know, for example, that in this limit Eq.\ref{eqn:strf} is not valid. 
\end{enumerate}

We find a significant dependence of the values of $\delta_c$ on the  field strength, and particularly in the physically interesting region around $f_{R0}=-10^{-6}$ -- currently close to the upper bound permissible by observations or theoretical considerations. We observe a strong environmental dependence with a significant trend: when reducing the dispersion of the smoothing Gaussian (and thus approaching pure top-hat distribution) we deviate further away from the analytical prediction in \cite{Schmidt}. This result shows that the non-linear chameleon properties of the $f(R)$ models strongly affect its behavior; thus analytical approximations based on linear predictions should be viewed as simple guidelines. However our error bars are still large, a more careful study is needed to make any definitive statements about $\delta_c$. 

Studying smoothed top-hat initial profiles and obtaining estimates for $\delta_c$ is the first step in studying spherical halos in $f(R)$ gravity. Further work is needed in understanding realistic halos with differing masses and environment. It would also be interesting in future work to study the abundance and clustering properties of halos: mass function and halo bias. 

\bigskip

{\it Acknowledgments:}
We are grateful to Wayne Hu, Mike Jarvis, Justin Khoury, 
Matt Martino, and Ravi Sheth for helpful discussions. We are grateful to Fabian Schmidt for comments on the manuscript and to him and Lucas Lombriser for sharing the results of ongoing work on related questions. This work supported in part by NSF grant AST-0607667. 

\section*{APPENDIX A: RELAXATION SCHEME FOR SOLVING THE SYSTEM OF NONLINEAR ODES}

As discussed in  ~\cite{HuSaw} (p.8) the primary equation we need to solve (Eq.\ref{eqn:fr1} and subsequently Eq.\ref{eqn:SODE1} and Eq.\ref{eqn:SODE2}) is non-linear and cannot be solved as an initial value problem as the homogeneous equation has exponentially growing and decaying Yukawa solutions $e^{(\pm \alpha r)}/r$. Initial-value integrators have numerical errors that would stimulate the positive exponential, whereas relaxation methods avoid this problem by enforcing the outer boundary at every step. So we also employ a relaxation method for solving 2-point boundary problems in ODE. We employ a Newton's method \cite{NRC} with dynamical allocation of the mesh grid. The mesh allocation function is taken to be logarithmic with its higher density at the origin, which is the primary region of interest and where we expect the solution to be more rapidly changing. As a guess solution for each step we utilize the relaxed solution of the previous step, while for the initial guess at the beginning of the simulation we use a linear solution. Generally if we have a system of discretized first order ordinary differential equations in the form: 
\begin{equation}
0 = \mathbf{E}_k \equiv \mathbf{y}_k - \mathbf{y}_{k-1} - (x_k - x_{k-1})\mathbf{g}_k (x_k, x_{k-1}, \mathbf{y}_k, \mathbf{y}_{k-1}) \quad k=2..M,
\end{equation}
 where the index $k$ spans the number of grid points $2..M$, the vector $\mathbf{E}$ consists of the system of $N$ discretized 1st order ODEs at each point (and has a total of $N*M$ components - $N*(M-1)$ from differential equations and $N$ from boundary conditions). $\mathbf{E}_1$ and $\mathbf{E}_{M+1}$ describe the boundary conditions. 
So a Taylor expansion with respect to small changes $\Delta {\mathbf y}_k$ looks like:
\begin{align}
&{\mathbf E}_k ( {\mathbf y}_k +\Delta {\mathbf y}_k, \mathbf{y}_{k-1} + \Delta \mathbf{y}_{k-1} ) \approx \\ &\approx {\mathbf E}_k( {\mathbf y}_k, \mathbf{y}_{k-1} ) + \sum_{n=1}^{N} \frac{\partial {\mathbf E}_k}{\partial y_{n,k-1}} \Delta y_{n,k-1} + \sum_{n=1}^{N} \frac{\partial {\mathbf E}_k}{\partial y_{n,k}} \Delta y_{n,k}, \nonumber
\end{align}
For a solution we want the updated value ${\mathbf E}_k ( {\mathbf y}_k +\Delta {\mathbf y}_k, \mathbf{y}_{k-1} + \Delta \mathbf{y}_{k-1} )$ to be zero, which sets up a matrix equation:
\begin{equation}
\sum_{n=1}^{N} S_{j,n} \Delta y_{n,k-1} + \sum_{n=N+1}^{2N} S_{j,n} \Delta y_{n,k} = -E_{j,k},
\end{equation}
where
\begin{equation}
 S_{j,n} = \frac{\partial E_{j,k}}{\partial y_{n,k-1}}, \quad  S_{j,n+N} = \frac{\partial E_{j,k}}{\partial y_{n,k}},
\end{equation}
and the quantity $S_{j,n}$ is a $N\times2N$ matrix at each point. Analogously we obtain similar algebraic equations on the boundaries.
Considering our problem we look at Eqs. (\ref{eqn:SODE1} and \ref{eqn:SODE2}) to obtain (after discretization and using the appropriate variables):
\begin{eqnarray}
  E_{1,k}=(y_k -y_{k-1}) - (r_k - r_{k-1})\left(\frac{r_k + r_{k-1}}{2}\right) \frac{1}{\bar{f}_R} \frac{\Omega_{M,0}}{a \tilde c^2} *  \\  * \left[ a^3 \left( \frac{1}{a^3} +4 \frac{\Omega_{\Lambda,0}}{\Omega_{M,0}} \right) \left(\sqrt{\frac{r_k + r_{k-1}}{2}}e^{-(\frac{r_k + r_{k-1}}{4})}  -1 \right) - \delta (\frac{r_k + r_{k-1}}{2}) \right] \nonumber
\end{eqnarray}
\begin{equation}
 E_{2,k}=(u_k -u_{k-1}) - (r_k - r_{k-1}) \left(\frac{y_k + y_{k-1}}{2}\right) e^{-(\frac{r_k + r_{k-1}}{2})} 
\end{equation}
\begin{equation}
 S_{1,1,k} = -1, \quad S_{1,3,k} = 1 
\end{equation}
\begin{align}
  &S_{1,2,k} =  S_{1,4,k} = \left(\frac{r_k - r_{k-1}}{4}\right) \left(\frac{r_k + r_{k-1}}{2}\right)* \\ &*\left[ a^3 \left( \frac{1}{a^3} +4 \frac{\Omega_{\Lambda,0}}{\Omega_{M,0}} \right) \sqrt{\frac{r_k + r_{k-1}}{2}}e^{-(\frac{r_k + r_{k-1}}{4})}  \right] \nonumber
\end{align}
\begin{equation}
  S_{2,1,k} =  S_{2,3,k} = - \left( \frac{r_k - r_{k-1}}{2}\right) e^{-(\frac{r_k + r_{k-1}}{2})}
\end{equation}
\begin{eqnarray}
  S_{2,2,k} =\left( \frac{r_k - r_{k-1}}{2} \right) \left( \frac{y_k + y_{k-1}}{2} \right) e^{-(\frac{r_k - r_{k-1}}{2})}-1  \\   S_{2,4,k} =\left( \frac{r_k - r_{k-1}}{2} \right) \left( \frac{y_k + y_{k-1}}{2} \right) e^{-(\frac{r_k - r_{k-1}}{2})} +1
\end{eqnarray}
The boundary conditions are also easily translated in terms of the relaxation scheme.
\begin{equation}
 E_{2,0}= e^{u_1} - r_1 y_1, \quad E_{1,M+1}=e^{u_M} - r_M
\end{equation}
The notation here is probably a bit confusing. The quantity ${\mathbf E}$ is a vector which consists consecutively of 2 elements per $M-1$ grid points. In addition there is one element each at the beginning and the end that correspond to the boundary conditions. Thus the total length of ${\mathbf E}$ is $NM$, while the matrix $S$ has $NM\times NM$ elements. The task of relaxing the solution at each step of the scheme requires solving the matrix equation: 
\begin{equation}
\label{eqn:matrix}
S\cdot \mathbf{b} = {\mathbf E} 
\end{equation}
The vector $\mathbf{b}$ contains the updates $\Delta {\mathbf y}_k$. For a grid of 1000 points our equation requires a matrix of derivatives of size 2000x2000 elements. Fortunately it is sparsely populated and as such can be represented by a sparse array structure. This allows for the use of methods particularly designed for solving such systems, like Krylov's method, which we employ.

\section*{APPENDIX B: SPHERICAL COLLAPSE IN $\Lambda$CDM}

The evolution and collapse of spherical overdensities have been useful for modeling the formation of galaxy and
cluster halos. In GR the problem can be approached analytically and we will outline the derivation presented e.g. in ~\cite{Schmidt}. We start with the nonlinear and Euler equation for a non-relativistic pressureless fluid in comoving coordinates:
\begin{align}
 \frac{\partial \delta}{\partial t} + \frac{1}{a} \triangledown \cdot (1+\delta)\bf{v} &= 0 \nonumber \\ \frac{\partial \mathbf{v}}{\partial t} + \frac{1}{a} (\mathbf{v} \cdot \triangledown ) {\mathbf v} + H {\mathbf v} &= -\frac{1}{a} \triangledown \phi,
\end{align}
where $a(t)$ is the expansion scale factor, $H(t)=\dot{a}/a$, and $\phi$ is the ``Newtonian'' potential. These equations continue to be valid for modifications of gravity that remain a metric theory \cite{Peebles}. These can now be joined together to form a second order equation for $\delta$.
\begin{equation}
 \frac{\partial^2 \delta}{\partial t^2} + 2 H \frac{\partial \delta}{\partial t} - \frac{1}{a^2} \frac{\partial^2 (1+\delta) v^i v^j}{\partial x^i \partial x^j} = \frac{\triangledown \cdot (1+\delta) \triangledown \phi}{a^2}
\end{equation}
Solving this equation requires information about the velocity and potential fields. In the case of a spherical top-hat distribution,  to preserve the top-hat distribution, the velocity field must take the form ${\mathbf v} = A(t){\mathbf r}$ to have a spatially constant divergence. Its amplitude is related to the top-hat density perturbation through the continuity equation: 
\begin{equation} 
 \dot{\delta}+\frac{3}{a}(1+\delta)A =0
\end{equation}
This leads to:
\begin{equation}
 \frac{\partial^2 v^i v^j}{\partial x^i \partial x^j} = 12 A^2 = \frac{4}{3} a^2 \frac{\dot{\delta}^2}{(1+\delta)^2}. 
\end{equation}
Substituting the above relation, the equation for the evolution of $\delta$ becomes:
\begin{equation}
 \frac{\partial^2 \delta}{\partial t^2} + 2 H \frac{\partial \delta}{\partial t} - \frac{4}{3} \frac{\dot{\delta}^2}{(1+\delta)} = \frac{(1+\delta)}{a^2} \triangledown^2 \phi,
\end{equation}
which is completed by the Poisson's equation for the potential:
\begin{equation}
 \triangledown^2 \phi = 4 \pi G a^2 \delta \rho_m.
\end{equation}
It is common to express spherical collapse through the evolution of the radius of the top-hat. For that we use mass conservation:
\begin{equation}
 M = \frac{4 \pi}{3} r^3 \bar{\rho}_m (1+\delta) = {\rm const.}
\end{equation}
to obtain the following relation:
\begin{equation}
 \frac{\ddot{r}}{r} = H^2 +\dot{H} - \frac{\triangledown^2 \phi}{3 a^2} 
\end{equation}
Expressing derivatives in terms of scale factor $' = d/d {\rm ln} a$, with the useful substitution:
\begin{equation}
 w = \frac{r}{r_i} - \frac{a}{a_i},
\end{equation}
 and using Poisson's equation we obtain:
\begin{equation}
 w'' + \frac{H'}{H} w' = -\frac{1}{2} \frac{\Omega_m a^{-3} -2 \Omega_{\Lambda}}{\Omega_m a^{-3} + \Omega_{\Lambda}}w  -\frac{1}{2} \frac{\Omega_m a^{-3}}{\Omega_m a^{-3} + \Omega_{\Lambda}}(\frac{a}{a_i}+w) \Delta,
\end{equation}
where
\begin{equation}
 \Delta = \left( \frac{1}{1+ w a_i / a} \right)^3 (1+\delta_i) -1.
\end{equation}
In these coordinates collapse occurs when $w=-\frac{a}{a_i}$.
The task of computing $\delta_c$ now reduces to the following: for a given $a_i$ find an initial overdensity $\delta_i$ such that the collapse occurs at $a=1$. Then using the linear growth factor in $\Lambda CDM$ (see for example ~\cite{Peebles}) we extrapolate $\delta_i$ to the present epoch to obtain $\delta_c$ as: 


\begin{equation}
 \delta_c(r=0) = A \int \frac{\hat{D}(k, a=1)}{\hat{D}(k, a_{in})} \hat{\delta}(k, a_{in}) e^{ikr} |_{r=0} dk
\end{equation}

{}
\end{document}